\documentclass[twocolumn]{webofc}

\usepackage[varg]{txfonts}   % Web of Conferences font
\usepackage{hyperref}
\usepackage{url}
\hypersetup{colorlinks=true,citecolor=blue,urlcolor=blue,linkcolor=blue}
\renewcommand{\vec}[1]{\mbox{\boldmath $#1$}}

\begin{document}
\title{Nuclear shape dynamics in low-energy heavy-ion reactions}

\author{\firstname{Kouichi} \lastname{Hagino}\inst{1,2,3}
}

\institute{
Department of Physics, Kyoto University, Kyoto 606-8502, Japan
\and
Institute for Liberal Arts and Sciences, Kyoto University, Kyoto 606-8501, Japan
\and
RIKEN Nishina Center for Accelerator-based Science, RIKEN, 
Wako 351-0198, Japan
          }

\abstract{
We discuss recent theoretical developments in low-energy 
heavy-ion reactions. To this end, we put emphasis on a viewpoint 
of probing nuclear shapes with heavy-ion reactions. 
We first discuss a single-channel problem with an optical 
potential model. 
We particularly discuss a microscopic modeling of the 
imaginary part of an optical potential as well as a visualization of quantum interference phenomena observed in heavy-ion elastic scattering. 
We then discuss multi-channel scattering problems, and demonstrate 
that heavy-ion fusion reactions at energies around the Coulomb barrier are sensitive to the shape of colliding nuclei, providing a powerful tool to probe nuclear shapes. We finally point out that relativistic heavy-ion collisions have  
large similarities to low-energy heavy-ion reactions in the context 
of nuclear shape dynamics. 
}
\maketitle
\section{Introduction}

A goal of low-energy nuclear physics is to understand atomic nuclei 
as quantum many-body systems of nucleons, that is, to understand 
several properties of nuclei in terms of nucleon degrees of freedom. 
These can be mainly divided into two categories: the first category 
concerns static properties of atomic nuclei, such as 
the mass, the size, the shape, and excitations of atomic nuclei, while the second 
category deals with their dynamical properties. These are covered by nuclear structure physics and nuclear reaction physics, respectively. 
Of course, there is no clear separation between these two. As a matter of fact, one of the most important 
roles of nuclear reactions is to serve as a tool to investigate nuclear 
structure. For instance, by knocking out a nucleon from a nucleus using nuclear reactions, one can learn single-particle natures 
of nuclei \cite{wakasa2017}. Three-body scattering, such as the $d$+$p$ reaction, has been utilized to investigate a three-nucleon 
interaction \cite{sekiguchi2014}. Furthermore, heavy-ion fusion reactions have been used as a standard tool to synthesize superheavy elements \cite{hofmann2000,oganessian2015}. 

Besides this aspect, nuclear reaction has another aspect. 
That is, the reaction dynamics itself is rich and complex, making it an interesting topics to study. Good examples for this include a formation of a compound nucleus 
with neutron induced reactions and its relation to quantum 
chaos \cite{weidenmuller2010},  
dynamics of the pre-equillibrium process towards a compound 
nucleus \cite{Hofmann2008}, 
and heavy-ion fusion reactions as quantum tunneling phenomena 
with many degrees of freedom \cite{balantekin1998,hagino2012}.

In this contribution, we shall discuss both of these aspects of  
nuclear reactions. 
For the latter aspect, we shall discuss quantum interference phenomena observed in nuclear reactions. We shall also introduce a novel 
method to visualize the quantum interferences by taking the Fourier 
transform of the scattering amplitude. 
For the former aspect, we shall mainly focus on extracting nuclear 
shapes from nuclear reactions. 
In passing, the richness of nuclear reaction dynamics originates 
from the fact that 
a nucleus is a composite system. If a nucleus was a point particle, only elastic scattering took place. However, in reality, because of the many-body nature of a nucleus, there exist rich reaction processes, including not only elastic scattering but also inelastic 
scattering, particle transfer reactions, and fusion reactions. These 
nuclear reaction processes are not independent with each other, but 
they significantly affect one another. 
For instance, the ground 
state properties of colliding nuclei, such as the shape of the nuclei as well as their excitations,  significantly affect 
those nuclear reaction processes \cite{balantekin1998,hagino2012}. 
Using this property, 
information on nuclear deformation has been succesfully extracted for several 
nuclei from low-energy heavy-ion fusion reactions \cite{dasgupta1998}. 
In recent years, a similar idea has been employed in relativistic heavy-ion collisions as well, 
providing an interesting intersection between low-energy and high-energy heavy-ion reactions \cite{jia2024,star2025}. 
We shall also discuss this topic in this contribution. 

\section{Single-channel problems}

\subsection{Microscopic modeling of the imaginary part of an optical potential}

Let us start with the simplest model of nuclear 
reactions, that is, 
single-channel scattering with an optical potential. 
An optical potential is a complex potential given by, 
\begin{equation}
    V_{\rm opt}(r)=V(r)-iW(r), 
    \label{eq:v_opt}
\end{equation}
where $r$ is 
the relative coordinate between two colliding nuclei.  
This model can discribe either elastic 
scattering or absorption processes. 
The imaginary part of the optical potential 
causes a loss of the flux, thus the absorption, which is originated from reaction 
processes other than elastic scattering, such as 
inelastic scattering, transfer reactions, and fusion 
reactions. In this approach, the dynamics after absorption is treated as a black box. 

On the other hand, there are several reaction processes in which the dynamics after the absorption is important. These include fusion of superheavy nuclei and fusion of two $^{12}$C nuclei. For the former, 
a re-separation process after the touching, referred to as quasi-fission, plays a crucial role \cite{hinde2021}. 
On the other hand, for the latter, the astrophysical $S$-factor for the $^{12}$C+$^{12}$C system 
exhibits several prominent resonance structures in contrast to 
those for the $^{12}$C+$^{13}$C system, which show a much smoother energy 
dependence \cite{zhang2020}. 
It was argued in Ref. \cite{jiang2013} that the different behaviors of fusion 
cross sections, and thus the astrophysical $S$-factors, in these systems are caused by different properties of the 
compound nuclei, $^{24}$Mg and $^{25}$Mg, indicating that 
the resonances are isolated in the $^{12}$C+$^{12}$C fusion reaction at low energies, while the $^{12}$C+$^{13}$C fusion reaction lies in the overlapping resonance regime. 

\begin{figure}
\centering
%\vspace*{1cm}       % Give the correct figure height in cm
\includegraphics[width=8cm,clip]{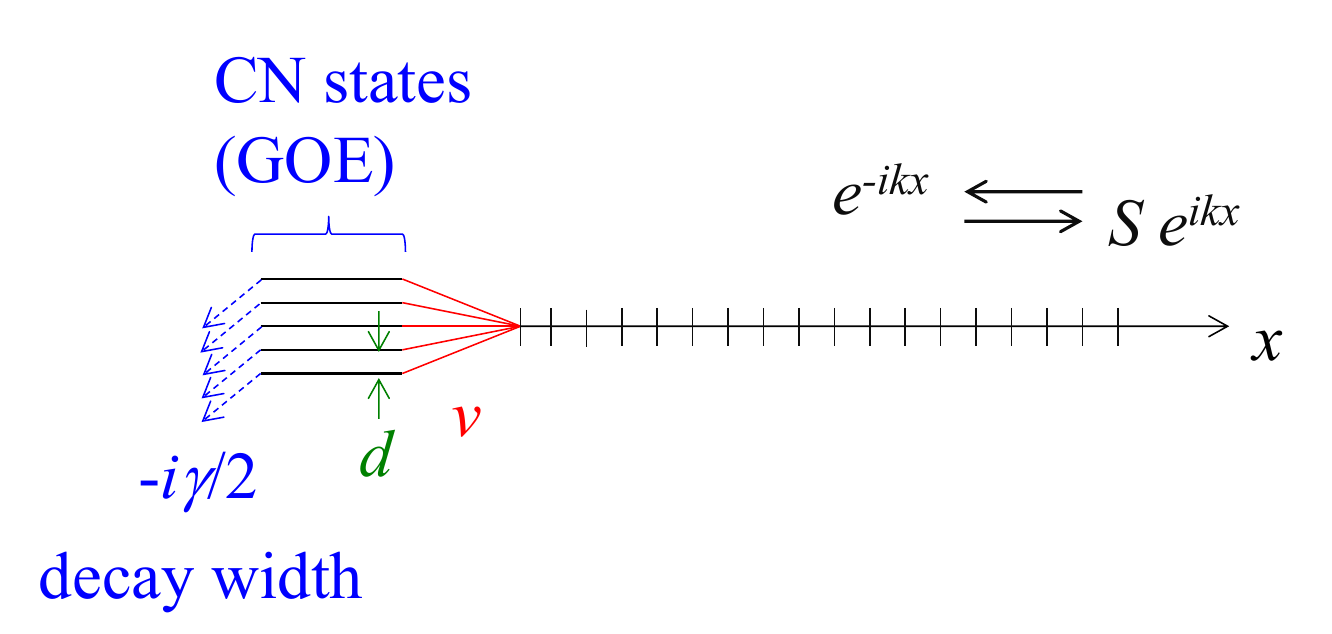}
\caption{
A schematic illustration of the schematic model proposed in Ref. \cite{hagino2025} 
for a microscopic modeling of an imaginary part of an optical potential. 
In this model, the entrance 
channel Hamiltonian in the discrete basis representation couples to compound nucleus states 
described with the random matrix based 
on the Gaussian Orthogonal Ensemble (GOE). 
$\gamma$ and $d$ are the decay width 
and the mean level spacing of the compound nucleus states, respectively. 
}
\label{fig:schematic}
\end{figure}

\begin{figure}
\centering
%\vspace*{1cm}       % Give the correct figure height in cm
\includegraphics[width=8cm,clip]{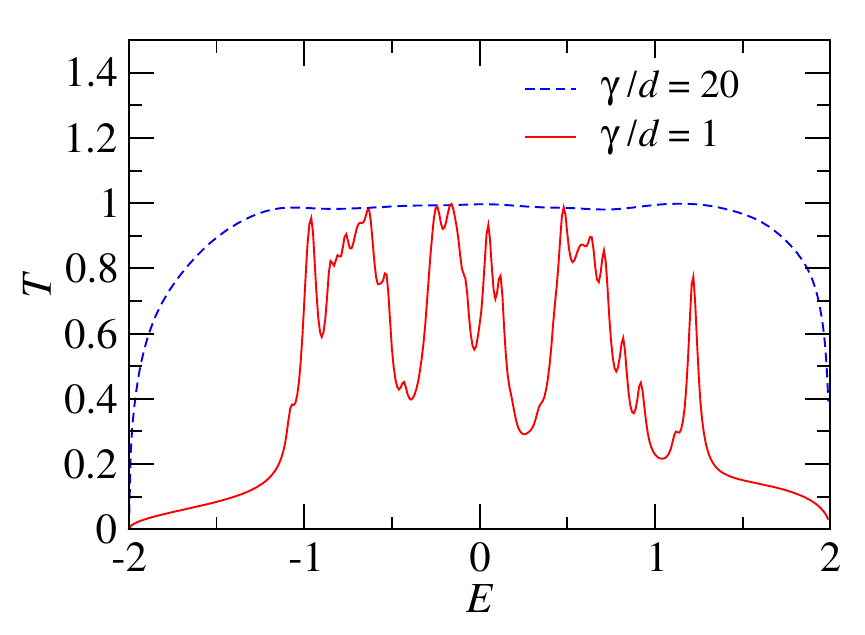}
\caption{Transmission coefficients $T$ as a function of the incident energy $E$ obtained with the 
schematic model introduced in Ref. \cite{hagino2025} (see Fig. \ref{fig:schematic}). 
The blue dashed line is for $\gamma/d=20$, while the red solid line for 
$\gamma/d=1$. 
}
\label{fig:schematic-transmission}
\end{figure}

To understand the underlying mechanism of the fusion reactions of the carbon isotopes, 
we recently constructed a simple schematic model, which consists of 
a random matrix Hamiltonian based on the Gaussian 
Orthogonal Ensemble (GOE) \cite{hagino2025}. 
See Fig. \ref{fig:schematic} for a schematic illustration of the model. 
Each of the GOE configurations posses a decay width $\gamma$ 
so that a part of the incident flux is absorbed 
by the GOE configurations. As a result, the absolute value of the $S$-matrix becomes 
less than one, $|S|<1$. This is how the imaginary part of the optical potential 
is modeled in this approach. By changing the interaction strength in the GOE Hamiltonian, 
one can control the level density $\rho$ of a "compound nucleus". 
Figure \ref{fig:schematic-transmission} shows the transmission coefficients for two different model parameters. 
The blue dashed line shows the transmission coefficient for $\gamma/d=20$, where 
$d=1/\rho$ is the mean level spacing. In this case, 
the 
transmission coefficients are close to unity, that is consistent 
with the strong absorption limit. 
On the other hand, the red solid line shows the transmission coefficients 
for $\gamma/d=1$, for which the transmission 
coefficients are much more structured as a function 
of $E$ with several prominent resonance peaks. 
It is interesting to observe that the transmission 
coefficients are close to unity at a few resonance 
energies, being consistent with the experimental observations in the C+C fusion reactions. 

\subsection{Imaging quantum interrference phenomena in heavy-ion elastic scattering}

One of the interesting aspects in low-energy heavy-ion reactions is that 
various quantum interference phenomena are seen in their differential cross sections. 
A textbook example for this is scattering of two identical particles, referred to as Mott 
scattering. For such scattering, 
a detector cannot distinguish scattering at angle $\theta$ from scattering at angle 
$\pi-\theta$, causing the characteristic interference pattern in the angular distribution \cite{bromley1961}. 
Another example is the nearside-farside 
interference \cite{hussein1984}, in which  
the nearside component corresponds to scattering with a positive 
impact parameter while the farside component is associated with a negative 
impact parameter. 
Notice that scattering takes place only at the edges  
of a nucleus due to a strong absorption inside a nucleus. 
The nearside and the farside components correspond to scattering at the opposite sides of the edges, 
and thus 
has a close analogy to scattering through two slits in the double slit problem \cite{hussein1984}. 

These quantum interference phenomena can be described with an optical potential, Eq. (\ref{eq:v_opt}). 
Using the optical potential, one can construct the scattering amplitude $f(\theta)$ e.g., with the partial wave decomposition technique, 
from which the differential cross sections $d\sigma/d\Omega$ can be computed as $d\sigma/d\Omega=|f(\theta)|^2$. 
By decomposing the scattering amplitude $f(\theta)$ into two or more components, 
$f(\theta)=\sum_if_i(\theta)$, one can obtain the interferences among the different 
components of the scattering amplitude. 

\begin{figure}
\centering
\includegraphics[width=8cm,clip]{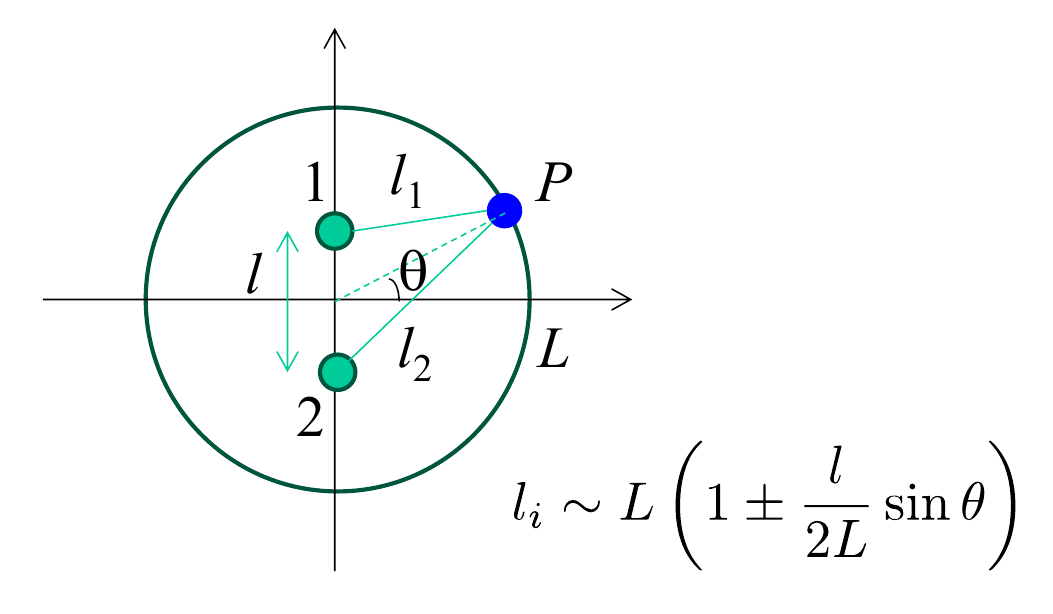}
\includegraphics[width=8cm,clip]{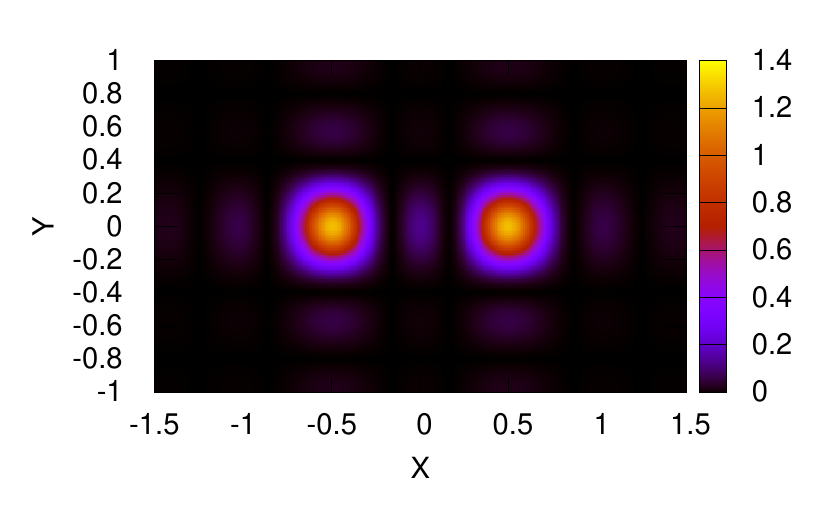}
\caption{(Upper panel) A schematic illustration of a setup of the double-slit problem. 
(Lower panel) The imaging of the double-slit problem with Eq. (\ref{eq:imaging}) on the two-dimensional $(X,Y)$ plane. 
The parameters are taken to be $l=1,k=30$, $\theta_0=0$, and $\Delta\theta=\Delta\varphi=$ 15 degree. 
}
\label{fig:double-slit}
\end{figure}

We would like to ask a question here: what would happen if one takes the Fourier transform of the scattering 
amplitude, $f(\theta)$? That is, 
\begin{equation}
  \Phi(X,Y)\propto \int^{\theta_0+\Delta\theta}
_{\theta0-\Delta\theta}d\theta\,
\int^{\Delta\varphi}
_{-\Delta\varphi}d\varphi\,e^{ik(\theta X+\varphi Y)}f(\theta). 
\label{eq:imaging}
\end{equation}
Let us first consider a double slit problem with classical waves 
with the amplitude $A$, the wave length $\lambda$, and the angular frequency $\omega$. For this problem,  
the total amplitude at an observation point P located at angle $\theta$ from the center of 
the two slits reads 
\begin{eqnarray}
    f(\theta)
   &=&A\sin\left(\frac{2\pi}{\lambda}l_1-\omega t\right)+A\sin\left(\frac{2\pi}{\lambda}l_2-\omega t\right), \\
   &\propto&\cos\left(\frac{\pi l}{\lambda}\sin\theta t\right), 
\end{eqnarray}
where 
$l_1$ and $l_2$ are the distance of the point P from each of the slits, and $l$ is the distance between 
the slits (see the upper panel of Fig. \ref{fig:double-slit}). 
Here, we have used the fact that $l$ is much smaller than the distance of P from the slits, $L$. 
Substituting this amplitude into Eq. (\ref{eq:imaging}), one obtains \cite{hashimoto2023}
\begin{eqnarray}
    \Phi(X,Y)&\propto& \sigma(kY\Delta\varphi)\left\{
    \sigma\left[\left(kX+\frac{kl}{2}\cos\theta_0\right)\Delta\theta\right]\right. \nonumber \\
    &&\left.+\sigma\left[\left(kX-\frac{kl}{2}\cos\theta_0\right)\Delta\theta\right]\right\},
\label{eq:imaging-doubleslit}
\end{eqnarray}
where $k=2\pi/\lambda$ is the wave length and the function $\sigma(x)$ is defined as $\sigma(x)=\sin x/x$. 
The lower panel of Fig. \ref{fig:double-slit} shows $|\Phi(X,Y)|^2$  
on the two-dimensional $(X,Y)$ plane for $l=1,k=30$, $\theta_0=0$, and $\Delta\theta=\Delta\varphi=$ 15 degree. 
One can see that this function 
shows a double peaked structure, and thus Eq. (\ref{eq:imaging})  
corresponds to making an image of the scattering waves onto a two-dimensional 
screen, whose coordinates are specified by $X$ and $Y$. 
Notice that the function $\sigma(x)$ is peaked at $x=0$, and thus Eq. (\ref{eq:imaging-doubleslit}) indicates that the 
peaks of $|\Phi(X,Y)|^2$ appear at $X=\pm\frac{kl}{2}\cos\theta_0$. This actually 
coincides with the peak position shown in Fig. \ref{fig:double-slit}. 
Notice that this method has been subsequently used to demonstrate that 
scattering of a string in the string theory is similar to a double slit problem \cite{hashimoto2023}. 

\begin{figure}
\centering
\includegraphics[width=8cm,clip]{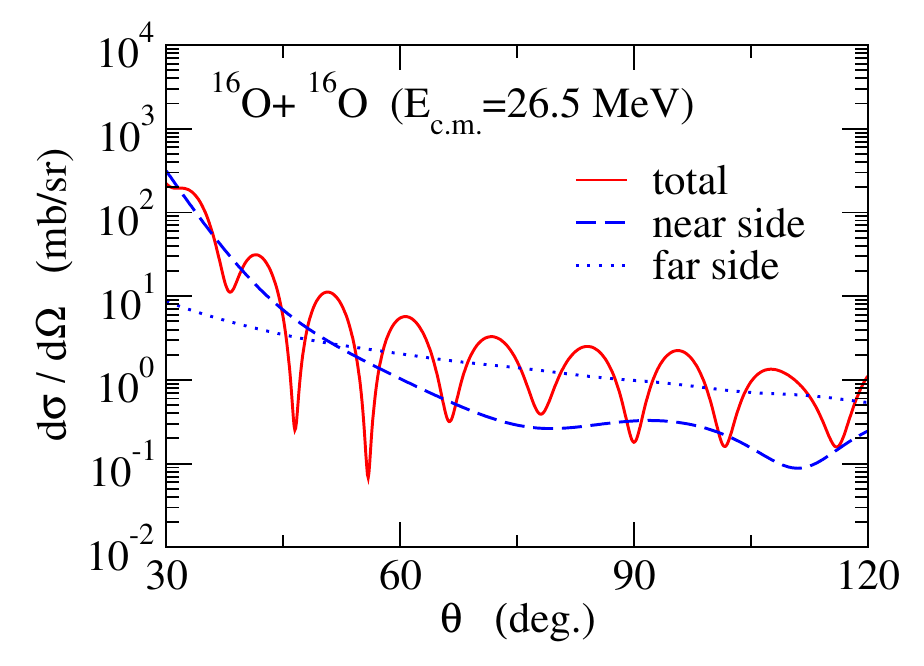}
\includegraphics[width=8cm,clip]{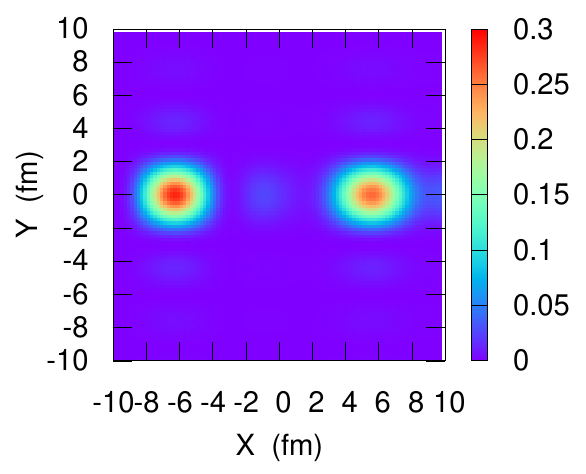}
\caption{
(Upper panel) Differential cross sections for 
$^{16}$O+$^{16}$O elastic 
scattering at $E_{\rm c.m.}$ = 26.5 MeV obtained with the optical potential model. 
The anti-symmetrization between the projectile and the target nuclei is ignored for simplicity. 
The solid line shows the result of the optical 
potential model calculation, while the dashed and the dotted lines show its decomposition 
to the near-side and the far-side components, respectively. 
(Lower panel) 
The visualization of $^{16}$O+$^{16}$O elastic 
scattering at $E_{\rm c.m.}$ = 26.5 MeV on a 
two-dimensional screen, whose coordinates are 
specified by $X$ and $Y$. The left and the right peaks correspond to the far-side and the near-side 
components, respectively. 
}
\label{imaging-16o16o}
\end{figure}

Let us next apply the same method to heavy-ion elastic scattering \cite{hagino2024,heo2025}. 
The upper panel of 
Fig. \ref{imaging-16o16o} shows differential cross sections for elastic scattering between two $^{16}$O nuclei 
at $E_{\rm c.m.}$ = 26.5 MeV. The solid line shows the result of an optical potential model 
calculation, while the dashed and the dotted lines show its decomposition into the near-side 
and the far-side components, respectively. 
See Ref. \cite{hagino2024} for the actual values of the parameters in the optical potential.  
For simplicity, we have ignored the symmetrization of the colliding nuclei. 
The imaging of the scattering, $|\Phi(X,Y)|^2$, 
obtained with $\theta_0=55$ deg., $\Delta\theta=\Delta\varphi$=15 deg. is shown in the lower panel of Fig. \ref{imaging-16o16o}. 
One can see a clear two-peaked structure, as in the double slit problem. 
By decomposing the scattering amplitude into the near-side and the far-side components, it can be 
shown that the peak at positive values of $X$ corresponds to the near-side component while the 
peak at negative values of $X$ corresponds to the far-side component \cite{hagino2024}. 

In this way, by visualizing heavy-ion reactions with the imaging technique 
presented here, one can gain an intuitive picture of quantum interference phenomena. 
An advantage of this method is that one can obtain an idea on how many different processes interfere with each 
other in a nuclear reaction process without explicitly decomposing the scattering amplitude. 
This property has been utilized in Ref. \cite{heo2025} to discuss the nearside-farside and the 
barrier-wave-internal-wave interferences in $\alpha$+$^{40}$Ca scattering. 

\section{Multi-channel problems}

\subsection{Subbarrier enhancement of fusion cross sections}

\begin{figure}
\centering
\includegraphics[width=8cm,clip]{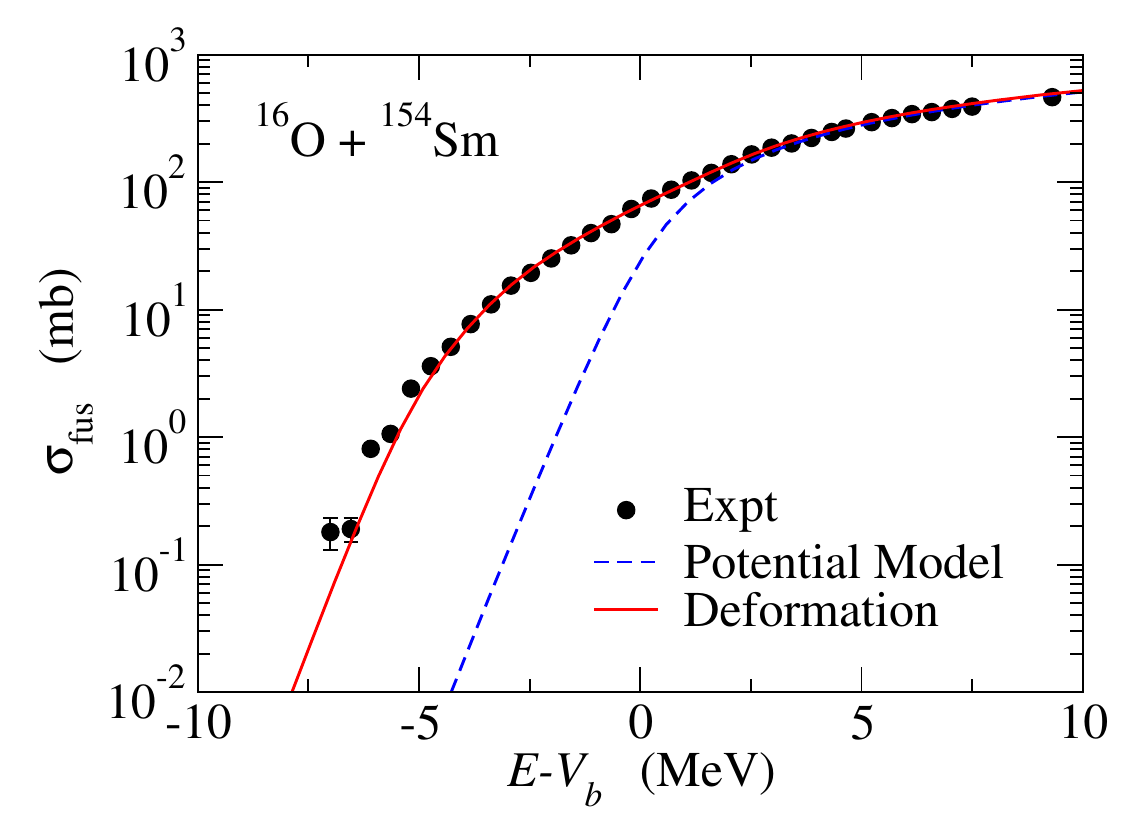}
\caption{
Fusion cross sections for the 
$^{16}$O+$^{154}$Sm system as a function of the 
incident energy in the center of mass frame, which 
is measured with respect to the height of the 
Coulomb barrier, $V_b$=60.35 MeV. 
The dashed line is obtained with a single-channel 
calculation, while the solid line takes into account 
the deformation effect of the $^{154}$Sm nucleus with 
the orientation average formula, Eq. (\ref{fusion}). 
The experimental data are taken from Ref. \cite{leigh1995}.
}
\label{fig:o+154sm}
\end{figure}

We next discuss multi-channel problems \cite{hagino2022}. We in particular discuss the role of nuclear deformation 
in heavy-ion fusion reactions at energies around the Coulomb barrier \cite{hagino2012,hagino2022,hagino1999}. 
A typical example is the $^{16}$O+$^{154}$Sm fusion reaction, whose cross sections are shown in Fig. \ref{fig:o+154sm} as a function 
of the incident energy in the center of mass frame. Here the energy is measured with respect to 
the height of the Coulomb barrier, $V_b$=60.35 MeV, which is estimated with a typical Woods-Saxon potential. 
In the figure, the dashed line is obtained with the potential model by assuming an inert $^{154}$Sm. This calculation 
largely underestimates the fusion cross sections at energies below the barrier. 
$^{154}$Sm is a typical deformed nucleus, and the large enhancement of fusion 
cross sections has been attributed to the deformation effect of $^{154}$Sm. 
In fact, by taking into account the deformation of $^{154}$Sm with the orientation average formula \cite{hagino2012,hagino2022}, 
\begin{equation}
\sigma_{\rm fus}(E)=\int^1_0d(\cos\theta)\,\sigma_{\rm fus}(E;\theta),
\label{fusion}
\end{equation}
where $\theta$ is the orientation angle of $^{154}$Sm with respect to the beam direction, and $\sigma_{\rm fus}(E;\theta)$ is the fusion 
cross section for a fixed orientation angle, 
one obtains the solid line, which well reproduces the experimental fusion cross sections. 

The orientation average formula, Eq. (\ref{fusion}), can be derived as follows. 
For an axially deformed target nucleus, the total Hamiltonian of the system reads,
\begin{equation}
    H=-\frac{\hbar^2}{2\mu}\vec{\nabla}^2+H_{\rm rot}+V(r,\theta),
    \label{Htot}
\end{equation}
where $\mu$ is the reduced mass, $r$ is the relative coordinate between the projectile and the target 
nuclei, and $V(r,\theta)$ is the angle dependent inter-nucleus potential. 
$H_{\rm rot}$ is the rotational energy of the target nucleus given by, 
\begin{equation}
H_{\rm rot}=\frac{\hat{\vec{I}}^2\hbar^2}{2{\cal J}}=
-\frac{\hbar^2}{2{\cal J}}\left(\frac{1}{\sin\theta}\frac{\partial}{\partial\theta}\left(\sin\theta
\frac{\partial}{\partial\theta}\right)+\frac{1}{\sin^2\theta}\frac{\partial^2}{\partial\varphi^2}\right),
\end{equation}
where $\hat{\vec{I}}$ is the angular momentum for the rotational motion of the deformed target nucleus, 
${\cal J}$ is the momentum inertia, and $\varphi$ is the azimuthal angle. 
For the medium heavy nuclei, such as $^{154}$Sm, the rotational energies 
are much smaller than the typical energy scale of the reactions, and the rotational Hamiltonian $H_{\rm rot}$ may be 
neglected in the total Hamiltonian, (\ref{Htot}). In this case, the total Hamiltonian is diagonal with respect to 
the angle $\theta$. Since a small rotational energy corresponds to a large moment of inertia ${\cal J}$, 
the angle $\theta$ is frozen during the reaction. Since the wave function for 
the ground state of the rotational motion is given by $Y_{00}(\theta)$, the total 
fusion cross section is given as a weighted average of fusion cross sections for a fixed value of $\theta$ 
with the weight factor given by $|Y_{00}(\theta)|^2$. This leads to the orientation average formula given by 
Eq. (\ref{fusion}). 

In this adiabatic limit, the rotational motion is so small that the initial configuration is retained during the 
reaction. Since the ground state is given as a linear superposition of a deformed wave function with 
different orientations, the orientation average formula is nothing but taking a snapshot of a deformed nucleus. 
Notice that this does not mean that the deformed nucleus is not rotationally excited during the reaction. The truth is 
rather opposite: the nucleus is coherently excited to all the rotational states. 
Notice that, from the uncertainty principle, one needs to superpose all 
the angular momentum states if one would like to freeze the orientation angle, that is, 
\begin{equation}
|\theta\rangle=\sum_{I=0}^\infty w(\theta)|I\rangle,~~~w(\theta)=\langle I|\theta\rangle.
\end{equation}
In fact, the orientation average formula, Eq. (\ref{fusion}), can be derived from the coupled-channels 
equations by neglecting the excitation energies of those states \cite{hagino2022,nagarajan1986}. 

\subsection{Barrier distributions}

In Ref. \cite{rowley1991}, Rowley, Satchler and Stelson proposed the so called fusion barrier distribution $D_{\rm fus}(E)$, that is defined as 
\begin{equation}
    D_{\rm fus}(E) = \frac{d^2(E\sigma_{\rm fus})}{dE^2}. 
    \label{bdistribution}
\end{equation}
For a single-channel problem, this function yields a Gaussian-like function centered at the 
barrier height energy \cite{hagino2012,dasgupta1998}. 
For a multi-channel problem, a single barrier is replaced by a multiple of barriers, and the barrier distribution (\ref{bdistribution}) 
provides a way to visualize how these barriers are distributed. 
For instance, in Eq. (\ref{fusion}), the barriers 
are specified for each value of $\theta$, and these are distributed with the weight factor given by $\sin\theta$. 

\begin{figure}
\centering
\includegraphics[width=8cm,clip]{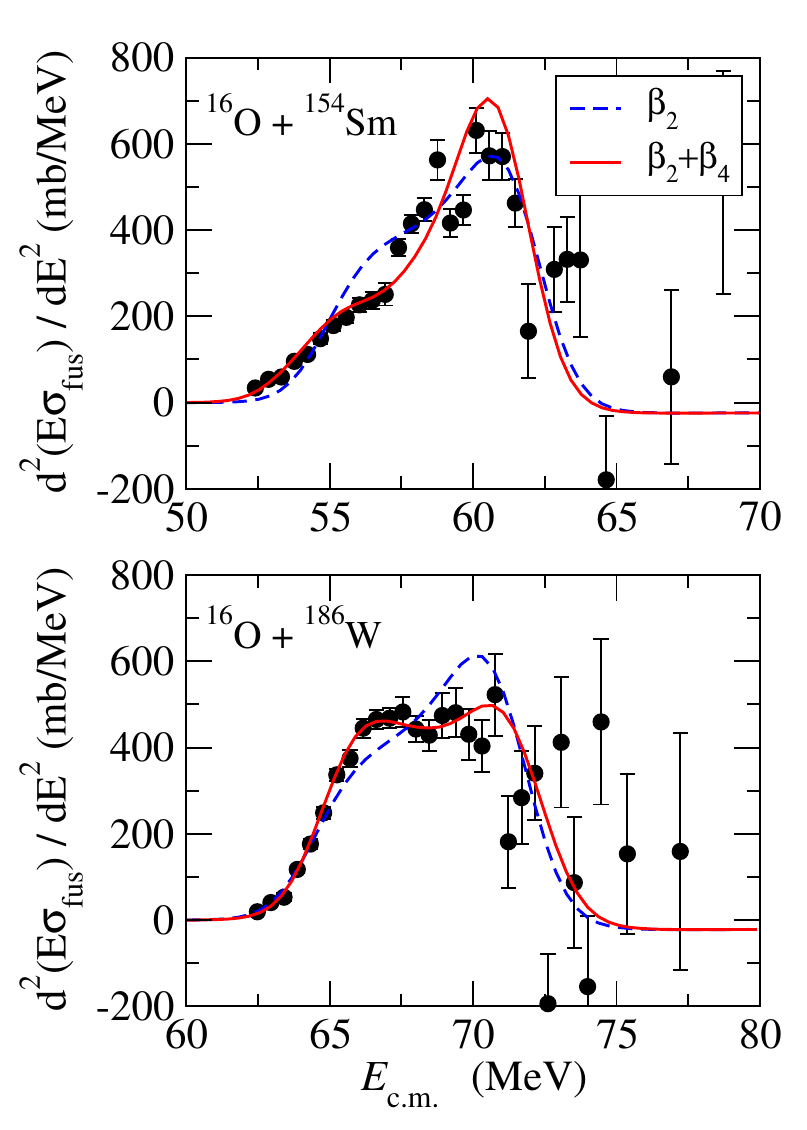}
\caption{
Fusion barrier distribution for the $^{16}$O+$^{154}$Sm system (the upper panel) 
and for the $^{16}$O+$^{186}$W system (the lower panel). The blue dashed lines 
take into account only the quadrupole deformation of the target nuclei, while the 
red solid lines takes into account in addition the hexadecapole deformation. 
The actual values of the deformation parameters are $(\beta_2,\beta_4)=(0.33,0.05)$ for $^{154}$Sm 
and $(\beta_2,\beta_4)=(0.29,-0.03)$ for $^{186}$W. 
The experimental data are taken from Ref. \cite{leigh1995}.
}
\label{fig:bd}
\end{figure}

It has been demonstrated that the barrier distribution is sensitive to nuclear structure, especially to the sign of hexadecapole 
deformation parameter \cite{dasgupta1998,leigh1995}. For instance, Fig. \ref{fig:bd} compares the barrier distribution for 
the $^{16}$O+$^{154}$Sm system with that for the $^{16}$O+$^{186}$W system. 
One can clearly see that the shape of the barrier distribution is significantly different for these two systems. 
It is known that both $^{154}$Sm and $^{186}$W have a similar quadrupole deformation parameter to each other. The 
dashed lines in the figure show the barrier distributions with the quadupole deformation parameter of each target nuclei.  
This calculation leads to similar barrier distributions to each other for these systems. 
The observed difference in the barrier distributions can be attributed to the hexadecapole deformation parameters, $\beta_4$. 
While $^{154}$Sm has a positive $\beta_4$ around +0.05, $^{186}$W has a negative $\beta_4$ around $-0.03$. 
The solid lines in Fig. \ref{fig:bd} are obtained by taking into account both the quadrupole and the hexadecapole deformation 
parameters of the target nuclei. One can now see that the different shapes of the barrier distributions can be well reproduced. 
It is interesting to notice that 
the hexadecapole deformation is by about 
one order of magnitude smaller than the quadrupole deformation, but still 
the shape of the barrier distribution is significantly affected by the hexadecapole deformation.  
In this way, the barrier distribution reveal how potential barriers are distributed, proving a fingerprint of reaction dynamics 
as well as a powerful tool to extract information on the nuclear shapes.  

\begin{figure}
\centering
\includegraphics[width=8cm,clip]{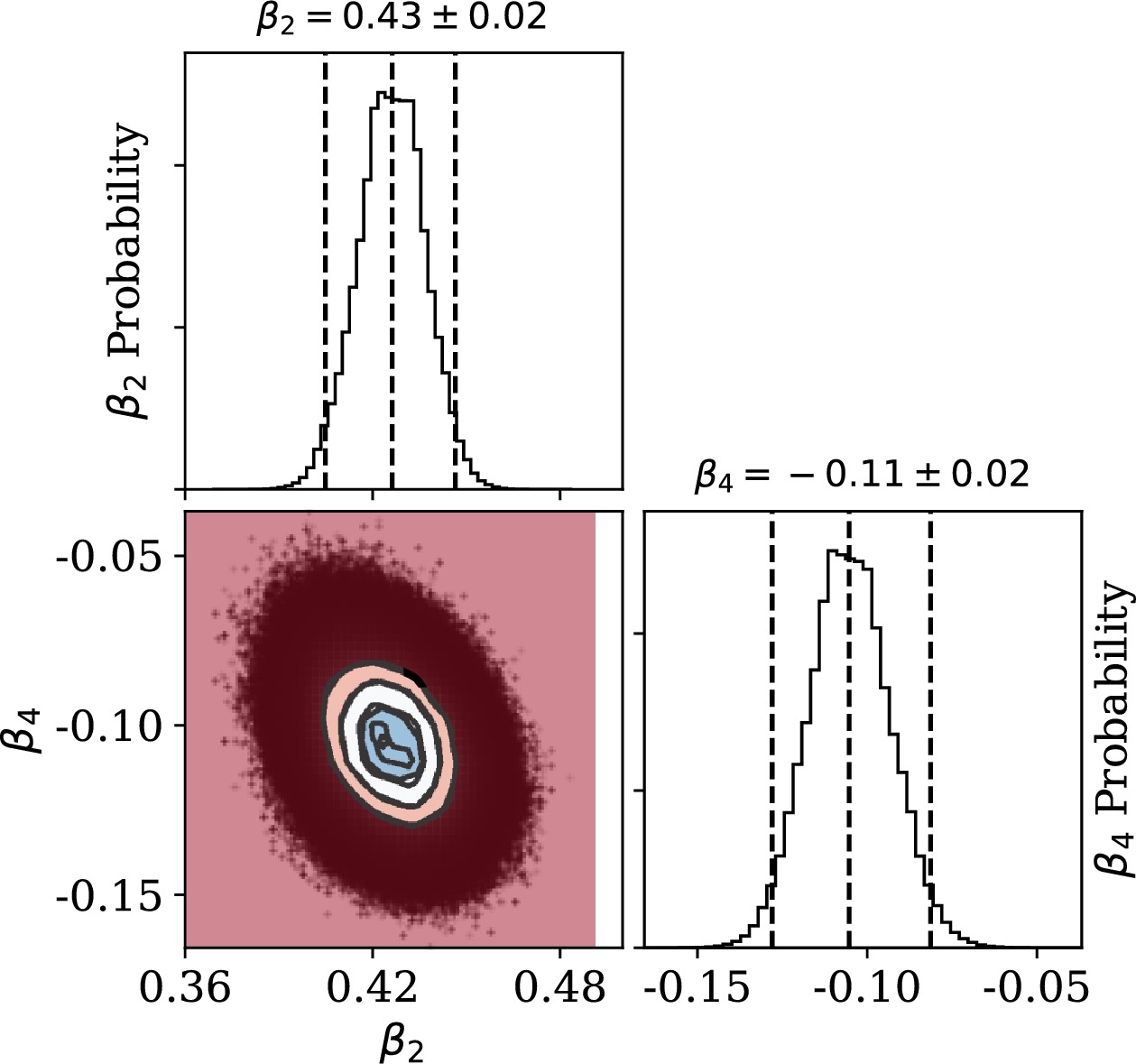}
\caption{
Multidimensional probability distributions of $\beta_2$ and $\beta_4$ for $^{24}$Mg obtained 
with a Bayesian analysis of the quasi-elastic barrier distribution for the 
$^{24}$Mg+$^{90}$Zr reaction. Taken from Ref. \cite{gupta2020}.
}
\label{fig:bd24mg}
\end{figure}

Using this property of the barrier distributions, the deformation parameters of $^{24}$Mg and $^{28}$Si have been 
recently extracted with high precision \cite{gupta2020,gupta2023}. To this end, 
Refs. \cite{gupta2020,gupta2023} analyzed the barrier distributions derived 
from quasi-elastic scattering \cite{timmers1995,hagino2004} using a Bayesian analysis 
with coupled-channels calculations (see Fig. \ref{fig:bd24mg}). 
The resultant values of the deformation parameters are 
$(\beta_2,\beta_4)=(0.43\pm0.02,-0.11\pm0.02)$ for $^{24}$Mg and 
$(\beta_2,\beta_4)=(-0.38\pm0.01,0.03\pm0.01)$ for $^{28}$Si, which are with much higher precision 
than the previous determinations, e.g., with proton inelastic scattering. 
This once again indicates that the barrier distribution offers a powerful method to probe nuclear shapes. 

\subsection{Emulator for multi-channel scattering}

In order to perform a Bayesian analysis presented in the last part of the previous subsection, 
one has to repeat coupled-channels calculations many times with different values of 
deformation parameters. 
This can be time-consuming if the number of channels included is large. 
In that situation, it is useful to have an emulator, which provides a good interpolation/extrapolation scheme 
to speed-up the calculations. 

\begin{figure}
\centering
\includegraphics[width=7cm,clip]{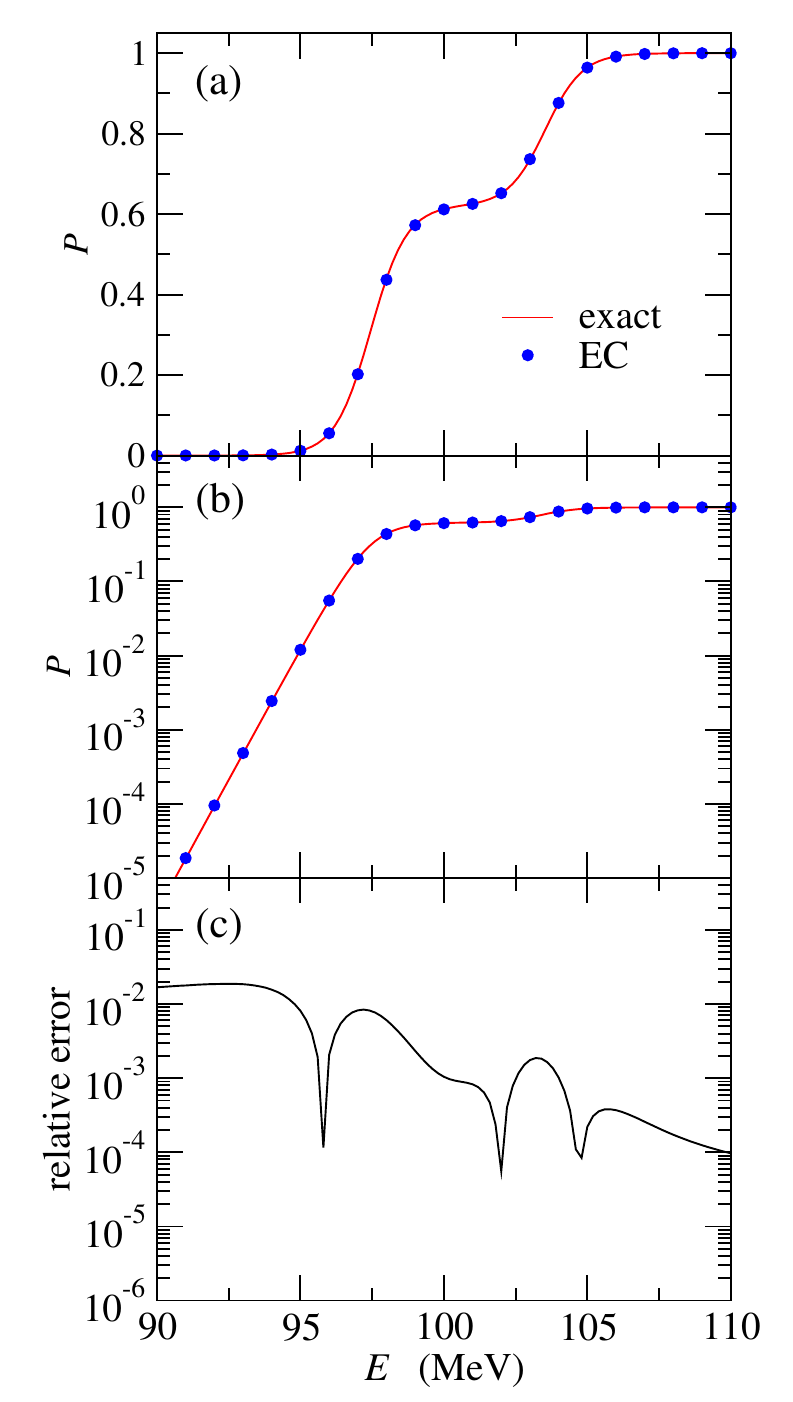}
\caption{
(a) The penetrability for 
a one-dimensional two-channel problem 
as a function of the energy $E$, plotted in the linear scale. 
The solid 
line shows the exact result, while the dashed line is obtained 
with the eigenvector continuation for the strength of the 
coupling potential. 
(b) The same as (a) but in the logarithmic scale. 
(c) The relative error defined by 
$|P_{\rm exact}-P_{\rm EC}|/P_{\rm exact}$, where $P_{\rm exact}$ and $P_{\rm EC}$ are the exact 
penetrability and that obtained with the eigenvector 
continuation, respectively. 
}
\label{fig:emulator}
\end{figure}

To construct an emulator, the eigenvector continuation (EC) 
has often been employed \cite{frame2018,duguet2024}. 
In this method, one linearly superposes a few eigenvectors of the model Hamiltonian with 
several different model parameters. That is, to solve the eigenvalue problem of a Hamiltonian $H(\theta)$, 
\begin{equation}
    H(\theta)|\Psi(\theta)\rangle=E(\theta)|\Psi(\theta)\rangle,
\end{equation}
where $\theta$ is a parameter in the Hamiltonian, one constructs an approximate solution by taking 
\begin{equation}
    \Psi(\theta)=\sum_{i=1}^Nc_i\Psi(\theta_i), 
\end{equation}
where $\Psi(\theta_i)$ is an eigenfunction of the Hamiltonian $H(\theta_i)$. Here, $N$ is the number of the basis functions, 
and the coefficients $\{c_i\}$ 
are determined variationally. 
This method was firstly applied to bound-state problems of atomic nuclei, but it has recently been applied to nuclear 
reactions as well \cite{furnstahl2020,drischler2021,Liu2024}.

Figure \ref{fig:emulator} shows the penetrabilities  
for a one-dimensional two-channel Hamiltonian, 
\begin{equation}
    H=\left(\begin{matrix}
V(x) & F(x) \\
F(x) & V(x)+\epsilon 
\end{matrix}
\right), 
\end{equation}
with a Gaussian barrier $V(x)=V_0\,e^{-x^2/2s^2}$ and 
a Gaussian coupling form factor $F(x)=F_0\,e^{-x^2/2s_f^2}$. 
The solid lines show the exact results 
with $V_0=100$ MeV, $F_0=3$ MeV, $s=s_f=3$ fm, and $\epsilon=$ 1 MeV, 
while the 
dots show the results of the eigenvector continuation 
with $F_{0i}=$1.5, 2.0, 2.5, 3.5, and 4.5 MeV to simulate 
the solution for $F_0=3$ MeV \cite{Hagino2025-2}. 
To implement the eigenvector continuation, we use the discrete basis method 
combined with the Kohn variational 
principle \cite{Hagino2024-2}. 
One can clearly see that the eigenvector continuation well reproduces 
the exact results, both in the linear scale (the top panel) 
and in the logarithmic scale (the middle panel). 
In particular, the exponential energy dependence of the penetrability at energies well below the barrier is successfully 
reproduced.
The relative error is shown in the bottom panel, indicating that the eigenvector continuation reproduces the 
exact result within about $10^{-4}$ at energies above the barrier, 
even though the error increases to $O(10^{-2})$ at energies below 
the barrier. 

\section{Relativistic Heavy-ion Collisions}

Let us lastly discuss relativistic heavy-ion collisions, 
for which 
there have been increasing interests in recent years 
in connection to a probe for nuclear deformations \cite{jia2024,star2025}. 
An idea of this approach is that the time (the energy) scale of heavy-ion collisions is much smaller (larger) 
at relativistic energies 
than the time (the energy) scale of nuclear motions, such that one can take a snapshot of a nucleus by choosing the central 
collision region where 
the projectile and the target nuclei significantly overlap with each other. 
In other words, the adiabatic approximation works well in the relativistic heavy-ion collisions, 
and the initial nuclear configuration is frozen during a collision, 
as in heavy-ion fusion reactions of a deformed nucleus discussed in Sec. 3. 
Using this property, the recent STAR collaborations successfully extracted the deformation parameter of $^{238}$U to be  
$\beta_2=-0.286\pm0.025$ and $\gamma=8.7\pm4.8^\circ$ \cite{star2024}. 

One of the main interests in nuclear structure physics is how to distinguish a static deformation from a dynamical deformation, that is, surface vibrations of {\it spherical} nuclei.  
In this connection, we mention that sub-barrier fusion reaction is significantly affected not only by a static deformation of 
deformed nuclei but also by a dynamical deformation 
of spherical nuclei \cite{esbensen1981}. Moreover they lead to considerably different 
fusion barrier distributions from each other \cite{hagino2012,hagino2022,dasgupta1998}. 
It would be interesting if this is the case also for relativistic heavy-ion 
collisions. 
To examine this, we particularly discuss the eccentricity parameter of the 
density distribution for the initial condition 
of relativistic heavy-ion collisions. 
Here, the eccentricity parameter is defined with the polar coordinate $(r,\theta,\phi)$ as
\begin{equation}
\epsilon_n\equiv 
%-\frac{\int d\vec{r}_\perp\,r_\perp^ne^{in\phi}\rho^{(z)}(\vec{r}_\perp)}
%{\int d\vec{r}_\perp\,r_\perp^n\rho^{(z)}(\vec{r}_\perp)}
%=
-\frac{\int d\vec{r}\,r^n\sin^n\theta e^{in\phi}\rho(\vec{r})}
{\int d\vec{r}\,r^n\sin^n\theta\rho(\vec{r})} 
=-\frac{\langle (x-iy)^n\rangle}{\langle (x^2+y^2)^{n/2} \rangle}.  \label{eq:eccentricity}
\end{equation}
To this end, we take 
the beam direction of a collision for the $z$-axis in the space-fixed coordinate system 
and assume for simplicity a central collision such 
that the projectile and the target nuclei 
have a complete overlap. 
It has been well recognized that the eccentricity parameter $\epsilon_n$ is directly 
correlated with the final-state anisotropic flows, $v_n$, for ultra-central collisions \cite{star2024,star2025}. 

To compute the eccentricity parameter (\ref{eq:eccentricity}), 
we shall employ 
a deformed Woods-Saxon density for $\rho(\vec{r})$ given by 
\begin{equation}
    \rho(\vec{r})=\frac{\rho_0}{1+e^{(r-R(\theta,\phi))/a}}, 
    \label{eq:Woods-Saxon}
\end{equation}
where $\rho_0$ and $a$ are the central density and the diffuseness parameter, respectively.  
Here,  $R(\theta,\phi)$ is the angle dependent radius given by 
\begin{equation}
     R(\theta,\phi)=R_0\left(1-\frac{1}{4\pi}\sum_{\lambda,\mu}|\alpha_{\lambda\mu}|^2
     +\sum_{\lambda,\mu}\alpha_{\lambda\mu}Y^*_{\lambda\mu}(\hat{\vec{r}})\right),
     \label{eq:radius}
\end{equation}
where $R_0$ is the radius parameter. In this equation, the second term in the parenthesis is due to the 
volume conservation, and $\alpha_{\lambda\mu}$ is the deformation parameter in the space-fixed system  
satisfying the condition $\alpha^*_{\lambda\mu}=(-1)^\mu\alpha_{\lambda,-\mu}$ \cite{RingSchuck}. 

In the harmonic oscillator model for spherical nuclei, the deformation parameter $\alpha_{\lambda\mu}$ acts 
as a coordinate of the harmonic oscillator. 
The classical Hamiltonian for the vibration reads 
\begin{equation}
    H=\frac{1}{2}\sum_{\lambda,\mu}\left(B_\lambda|\dot{\alpha}_{\lambda\mu}|^2+C_\lambda|\alpha_{\lambda\mu}|^2\right), 
    \label{eq:Hcl}
\end{equation}
where the dot denotes the time derivative, and 
$B_\lambda$ and $C_\lambda$ are the inertia and the stiffness parameters of the oscillator, respectively. 
Notice that, for a given $\lambda$, there exist $2\lambda+1$ independent harmonic oscillators with 
$\{\alpha_{\lambda\mu}\}$ as the coordinates. 
In quantum mechanics, the ground state of these harmonic oscillators has the zero point fluctuation, with 
the probability distribution of the coordinate given by, 
\begin{equation}
    P(x)=\frac{1}{\sqrt{2\pi\sigma_\lambda^2}}\,e^{-x^2/(2\sigma_\lambda^2)}, 
    \label{eq:ho-prob}
\end{equation}
where $\sigma_\lambda$ is the amplitude of the zero point motion, given by $\sigma_\lambda=\sqrt{\hbar/(2B_\lambda\omega_\lambda)}$ with 
$\omega_\lambda=\sqrt{C_\lambda/B_\lambda}$. 
The expectation value of the absolute square of the eccentricity parameter is then evaluated as, 
\begin{equation}
    \langle |\epsilon_n|^2\rangle 
    =\int \left(\prod_{\lambda,\mu} d\alpha_{\lambda\mu}\,P(\alpha_{\lambda\mu})\right)|\epsilon_n(\{\alpha_{\lambda\mu}\})|^2,
    \label{eq:ecc_vib}
\end{equation}
where $\epsilon_n(\{\alpha_{\lambda\mu}\})$ is the eccentricity parameter (\ref{eq:eccentricity}) for a given set of $\{\alpha_{\lambda\mu}\}$. 
With the probability distribution given by Eq. (\ref{eq:ho-prob}),  
the expectation value of 
$\sum_\lambda|\alpha_{\lambda\mu}|^2$ is computed as 
\begin{equation}
    \left\langle\sum_\mu|\alpha_{\lambda\mu}|^2\right\rangle
    =(2\lambda+1)\sigma_\lambda^2\equiv(\beta_\lambda)^2. 
    \label{eq:beta}
\end{equation}
In low-energy heavy-ion fusion reactions, the square root of this quantity is often referred to as the (dynamical) deformation 
parameter and is denoted by $\beta_\lambda$ \cite{hagino2012,hagino2022}. 
The value of the dynamical deformation parameter can be estimated from a measured electronic transition 
probability $B(E\lambda)\uparrow$ from the ground state of a spherical nucleus to an excited state as \cite{hagino2012,hagino2022},
\begin{equation}
    \beta_\lambda=\frac{4\pi}{3ZR_0^\lambda}\,\sqrt{\frac{B(E\lambda)\uparrow}{e^2}}, 
\end{equation}
where $Z$ is the atomic number of the nucleus. 

On the other hand, in the case of static deformation, 
one usually 
transforms the coordinate system to the body-fixed system with an appropriate choice of principle axes. 
For an axially-symmetric shape, this leads to \cite{hagino2025-3}
\begin{equation}
    \alpha_{\lambda\mu}=\beta_\lambda D^\lambda_{0\mu}(\Omega), 
     \label{eq:defpara-axial}
\end{equation}
where $D^\lambda_{\mu'\mu}$ is the Wigner D function and $\Omega$ is the Euler angle for 
the transformation from the body-fixed to the space-fixed frames. 
By substituting this expression into Eq. (\ref{eq:radius}), one can evaluate the eccentricity 
parameter $\epsilon_n(\Omega)$ for a given Euler angle $\Omega$. 
The average value of the absolute square of the eccentricity parameter reads,  
\begin{equation}
    \langle |\epsilon_n|^2\rangle 
    =\int \frac{d\Omega}{8\pi^2}\, |\epsilon_n(\Omega)|^2.
        \label{eq:ecc_rot}
\end{equation}

\begin{figure}
\centering
\includegraphics[width=8cm,clip]{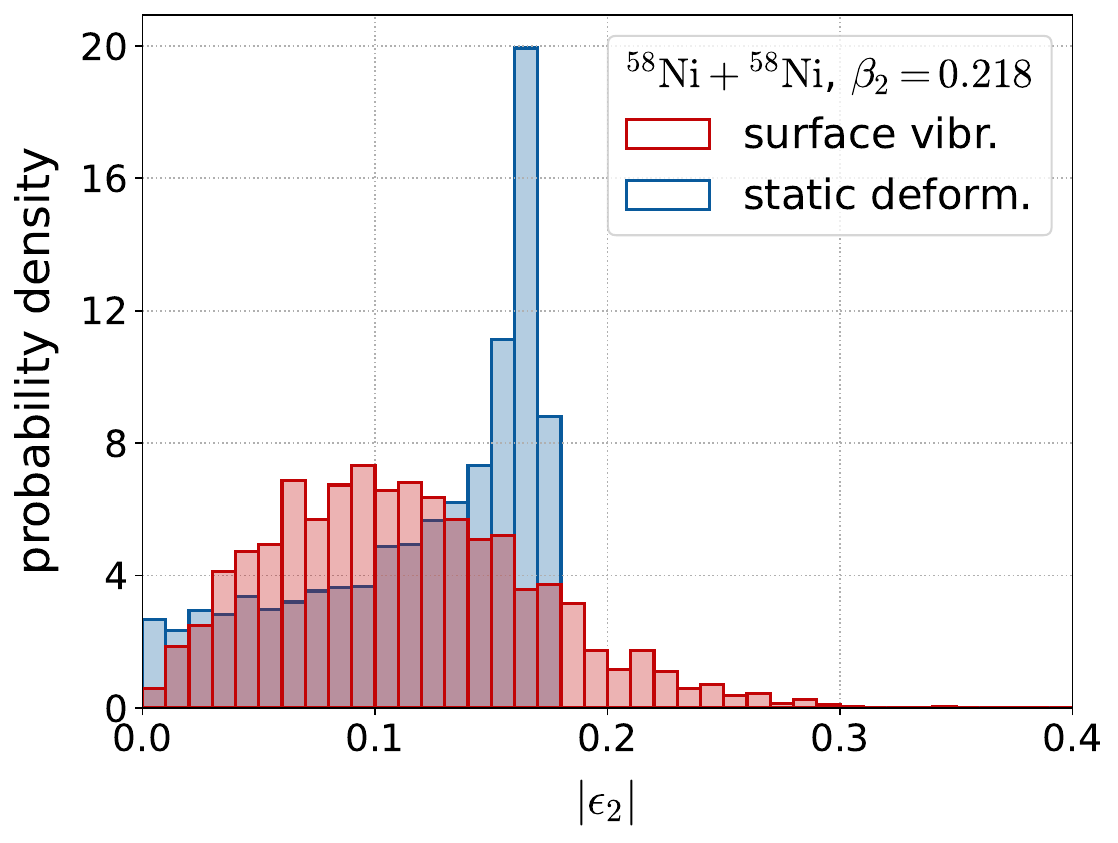}
\caption{
Probability densities of the 
quadrupole eccentricity parameter $|\epsilon_2|$ for the $^{58}$Ni+$^{58}$Ni collision. 
The red and blue colors denote the results for the surface vibration and the static deformation, respectively. 
}
\label{fig:dist58ni}
\end{figure}

\begin{figure}
\centering
\includegraphics[width=8cm,clip]{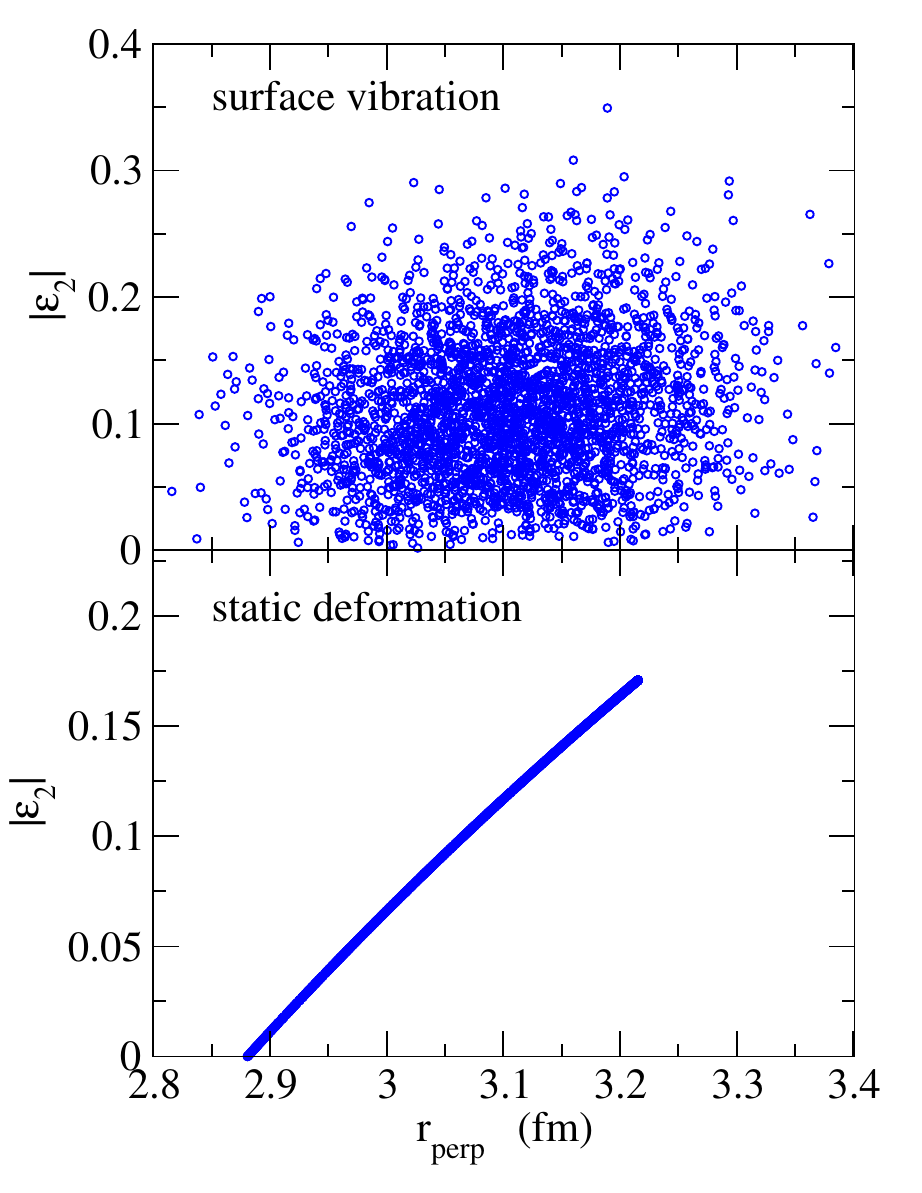}
\caption{
The distributions of the quadrupole eccentricity parameters $\epsilon_2$ for the 
the $^{58}$Ni+$^{58}$Ni collision 
plotted as a function of  
the root mean square radius on the perpendicular plane, 
$r_\perp\equiv\sqrt{\langle x^2+y^2\rangle}$. 
}
\label{fig:dist58ni-2}
\end{figure}

Figure \ref{fig:dist58ni} compares 
the distribution of the eccentricity parameters 
for the surface vibration (SV) with that for 
the static deformation (SD) for $^{58}$Ni+$^{58}$Ni collision. 
To this end, 
we generate $\{\alpha_{\lambda\mu}\}$ randomly according to the probability distribution in Eq. (\ref{eq:ho-prob}) 
for SV, while for SD only $\Omega$ is randomly sampled with fixed $\beta_\lambda$ for the axial shape.
The value of $\beta_2$ is estimated to be 0.218 from 
the measured $B(E2)$ for the transition from 
the ground state to the first 2$^+$ state at 1.45 MeV in 
$^{58}$Ni. 
One can clearly see in the figure that 
SV and SD lead to significantly 
different distributions of the eccentricity parameters from each other, despite that the mean values are similar 
to each other, that is, $\langle|\epsilon_2|\rangle$ is 0.112 and 0.119 
for SV and SD, respectively. 
This becomes even clearer if the eccentricity parameter for each sample is plotted as a function of 
the root mean square radius on the perpendicular plane, 
$r_\perp\equiv\sqrt{\langle x^2+y^2\rangle}$, as is shown in 
Fig. \ref{fig:dist58ni-2}. 

These suggest that relativistic heavy-ion collisions provide an interesting way to distinguish between dynamical deformations (i.e., surface fluctuations) and static deformations of atomic nuclei. 

\section{Summary}

We have discussed the dynamics of low-energy 
heavy-ion reactions. We first used the optical potential model and 
demonstrated that a part of the imaginary part of an optical potential 
can be well modeled with a simple model that consists of a random matrix for a compound nucleus. 
By changing the level density and the decay width of the compound nucleus states, 
the model can describe both the 
overlapping resonance regime and the isolated 
resonance regime, as in the fusion cross sections of 
the $^{12}$C+$^{12}$C and $^{12}$C+$^{13}$C systems. 
We next introduced the imaging technique for heavy-ion elastic scattering. 
This is to take the Fourier transform of the scattering amplitude. We have shown that 
this yields an intuitive view of quantum interference phenomena in heavy-ion reactions, 
such as a nearside-farside interference and a barrier-wave-internal-wave interference. 

We then discussed multi-channel problems, putting some emphasis on heavy-ion fusion 
reactions at subbarrier energies. It has been well known that fusion cross sections 
are largely enhanced at these energies due to collective excitations of colliding nuclei, 
such as the rotational excitations of a deformed nucleus and the surface vibrations of a 
spherical nucleus. Using fusion barrier distributions, the deformation parameters have 
been successfully extracted for several nuclei. For such analyses, one needs to carry out coupled-channels 
calculations many times for different values of deformation parameters. We 
have demonstrated that an emulator for multi-channel scattering provides 
a powerful tool for that purpose. 

Lastly, we have discussed relativistic heavy-ion collisions from the viewpoint of probing 
nuclear shapes. By computing the eccentricity parameters, we have demonstrated 
that a surface vibration and a static deformation lead to considerably different distributions 
of the eccentricity parameters. This suggests that relativistic heavy-ion collisions may provide   
a promising tool to distinguish between a static deformation and a dynamical deformation, making  
a good intersection between low-energy and high-energy nuclear physics. 

All these studies strongly demonstrate that nuclear reaction dynamics is rich and intriguing to study. 
This all comes from the fact that atomic nuclei are composite quatum many-body 
systems and, as a consequence, there is a strong interplay between nuclear structure and nuclear reactions. 
This has been known well in low-energy nuclear reactions, but recent studies 
on relativistic heavy-ion collisions have revealed that this is the case also in high-energy nuclear 
collisions. We are now at an interesting stage to further explore nuclear reaction dynamics 
across a wide range of energy scales. 

\section*{Acknowledgment}
We thank G.F. Bertsch, Y. Gupta, K. Heo, M. Kimura, 
M. Kitazawa, Z. Liao, K. Uzawa, T. Yoda, 
S. Yoshida for collaborations. 
This work was partly supported by JSPS KAKENHI Grant Number JP23K03414.

\bibliography{ref} 

\end{document}